\documentstyle[seceq,epsf]{style1}

\newcommand{\slsh}[1]{\mbox{$\not\! #1$}}
\newcommand{\bm}[1]{\mbox{\boldmath $#1$}}
\newcommand{\st}{{T}}

\newcommand{\bpt}{\bm p_T^{}}
\newcommand{\bkt}{\bm k_T^{}}
\newcommand{\bSt}{\bm S_T}
\newcommand{\ba}{\begin{eqnarray}}
\newcommand{\ea}{\end{eqnarray}}
\newcommand{\beq}{\begin{equation}}
\newcommand{\eeq}{\end{equation}}

\newcommand{\text}{\hbox}

\newcommand{\simorder}{\raisebox{-4pt}{$\, \stackrel{\textstyle >}{\sim} \,$}}
\newcommand{\simordertwo}{\raisebox{-4pt}{$\, \stackrel{\textstyle <}{\sim} \,$}}

\title{
Transverse polarization distribution and fragmentation functions
\footnote{Invited talk given at the Circum-Pan-Pacific RIKEN Symposium on ``High Energy Spin Physics'', RIKEN, Wako, Japan, November 3-6, 1999}
}

\author{
Dani\"el Boer
}

\inst{
RIKEN-BNL Research Center\\
Brookhaven National Laboratory, Upton, NY 11973, U.S.A.}

\recdate{
\today
}

\abst{
We discuss transverse polarization distribution and 
fragmentation functions, in particular, T-odd functions with 
transverse momentum dependence, which might be
relevant for the description of single transverse spin asymmetries. The
role of intrinsic tranverse momentum in 
the expansion in inverse powers of the hard scale 
is elaborated upon. The $\sin \phi$ single spin asymmetry in the
process $e \, \vec{p} \to e' \, \pi^+ \, X$ as recently reported by the 
HERMES Collaboration is investigated, in particular, by using the bag model.}

\begin{document}

\maketitle

\section{Introduction}
Transverse polarization distribution and 
fragmentation functions parameterize transverse spin effects in hard
scattering processes. Although in general little is known 
experimentally on most of these functions, some 
transverse spin experiments have been performed. For instance, 
large single transverse spin asymmetries have been observed \cite{Adams} 
in the process $p \, p^{\uparrow} \rightarrow \pi \, X$.
Therefore, the question arises which transverse polarization 
distribution and/or fragmentation functions are relevant for their description.
This might include so-called T-odd fragmentation functions, 
which are expected to arise due to final state interactions. 
The main observation is that the 
description of this specific process in terms of 
transverse spin functions will lead to power
suppressed single spin asymmetries, unless one takes into account the 
intrinsic transverse momentum of the partons, see e.g.\ Ref.\ \cite{Boer}. 
We will discuss how the transverse
momentum dependence of asymmetries can contain 
information (in terms of distribution and
fragmentation functions) at leading order, 
that would be power suppressed if integrated over
the transverse momentum.  

Especially the T-odd functions with transverse momentum dependence might be
relevant for the description of single transverse spin asymmetries, since
these functions link the transverse momentum and transverse spin (of either
quarks or hadrons) with a specific handedness. The different functions will
lead to different angular dependences. Hence, studying the angular 
dependences of asymmetries (and their transverse momentum dependence) 
is a most promising way to unravel the origin(s)
of transverse spin asymmetries. This is for instance demonstrated 
by a recent result by the 
HERMES Collaboration \cite{HERMES}, as will be discussed. 

\section{Distribution and fragmentation functions}

\begin{figure}[htb]
    \parbox{\halftext}{
                \epsfxsize=5cm \centerline{\epsfbox{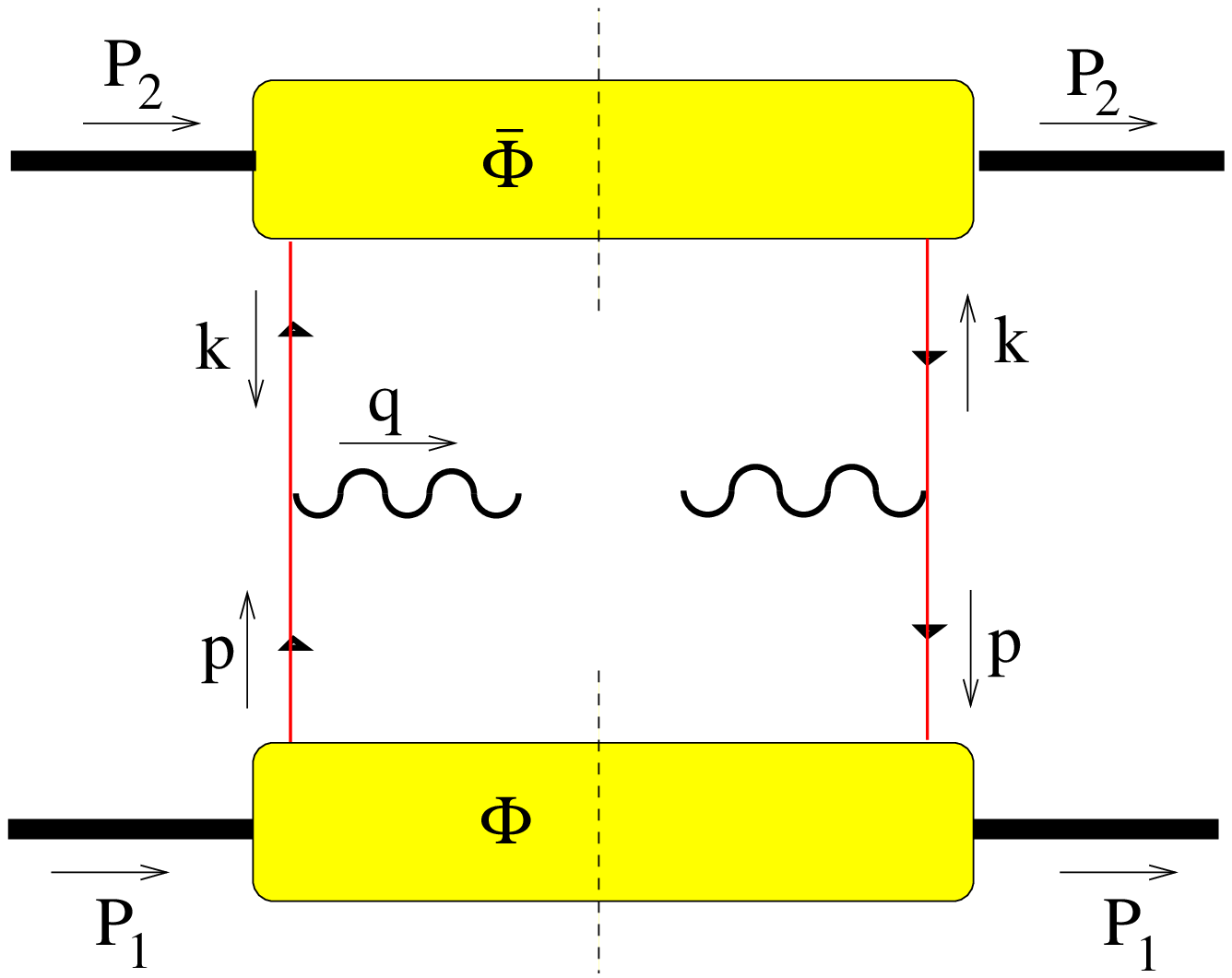}}
                \caption{\label{LODY}The leading order contribution to the
Drell-Yan process.}}
            \hspace{8mm}
            \parbox{\halftext}{
                \epsfxsize=5cm \centerline{\epsfbox{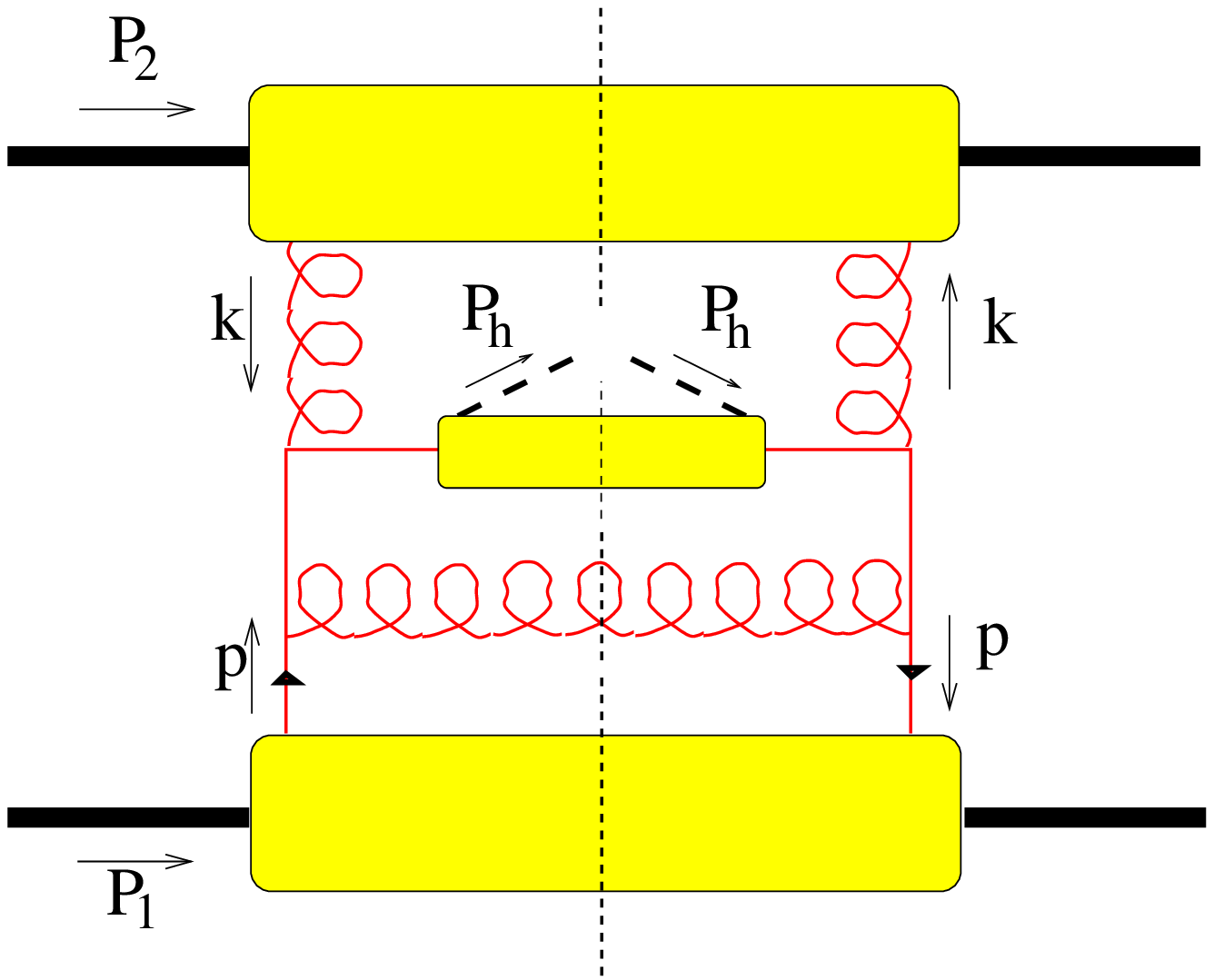}}
		\caption{\label{pppiX}A contribution to the
process $p \, p^{\uparrow} \rightarrow \pi \, X$.}}
\end{figure}

Transverse spin asymmetries in hadron-hadron collisions require an
explanation that involves quarks and gluons. A large scale allows for a 
factorization of such processes into parts describing the
soft physics convoluted with a hard subprocess cross section. 
We will first focus on the
Drell-Yan (DY) process, i.e.\ lepton pair production. 
In lowest order --the parton model approximation-- this process 
consists of two soft parts, the correlation functions called $\Phi$ and
$\overline \Phi$. In Fig.\ \ref{LODY}
the leading order diagram is depicted \cite{Ralst-S-79}. 
One can decompose the quark momenta
$p$ and $k$ into parts that are along the direction of the parent hadron,
the so-called lightcone momentum fractions, 
and deviations from that direction. In case one integrates over the transverse
momentum of the lepton pair one only has to consider the correlation 
functions as functions of the lightcone momentum fractions, for instance 
$\Phi(x)$. Its most general parameterization, that is in accordance with 
the required symmetries (hermiticity, parity, time reversal), is given by:
\beq
\Phi(x)=\frac{1}{2} \left[{f_1(x)} \mbox{$\not\! P\,$}+ 
{g_1(x)}\,\lambda\gamma_{5}\mbox{$\not\! P\,$}
+ {h_1(x)}\,\gamma_{5}\mbox{$\not\! S$}_{T}\mbox{$\not\! P\,$}\right].
\label{paramPhi}
\eeq
Other common notation is $q$ for $f_1$,  $\Delta q$ for $g_1$ and $\delta q $
or $\Delta_T q$ for $h_1$.

At the parton model level one finds the well-known {\em double\/} 
transverse spin 
asymmetry \cite{Ralst-S-79}, $A_{TT} \propto |\bm S_{1T}^{}|\;|\bm 
S_{2T}^{}|\;\cos(\phi_{S_1}+\phi_{S_2})\;
            h_1(x_1) \, \overline h_1(x_2)$, which is one possibility 
to obtain information on the transversity 
distribution function $h_1$, for instance at RHIC.
At this order there are no {\em single\/} transverse spin asymmetries. However,
these might arise from corrections to this lowest order diagram: 
the perturbative and/or the higher twist 
corrections. The former will typically yield\cite{Kane} 
single transverse spin asymmetries 
of the order $\alpha_s m_q/\sqrt{\hat s}$ which are expected to be 
small. The higher twist corrections require parameterizing 
the correlation function $\Phi(x)$ to
include contributions proportional to the hadronic scale (say the hadron
mass), 
\beq
\Phi(x)=\text{Eq.}\, (\ref{paramPhi}) + \frac{M}{2} 
\left[ {e(x)}\bm{1} + {g_T(x)} \gamma_5 \mbox{$\not\! S$}_{T}
+ {h_L(x)} \frac{\lambda}{2} \gamma_5 \left[\slsh{n_+},\slsh{n_-}\right]
\right]. 
\label{paramPhiM}
\eeq
The distribution functions {$e,g_T,h_L$} are so-called {twist-3} functions;
these will show up in the cross section suppressed by $M/Q$,
where $Q$ is a hard scale.  
At leading order in $\alpha_s$, i.e.\ $(\alpha_s)^0$, but at the order
$1/Q$, one 
finds \cite{Jaffe-Ji-91,Tangerman-Mulders-95a} no single or double
transverse spin asymmetries in the DY cross section. 
The functions $g_T$ and $h_L$ will only\footnote{The chiral-even, T-even 
function $g_T= g_1 + g_2$ has been studied 
by deep inelastic scattering of leptons off transversely 
polarized hadrons at SLAC and by SMC.} appear \cite{Jaffe-Ji-91} in 
the asymmetry $A_{LT}$.  

\begin{figure}[htb]
\epsfxsize=9cm \centerline{\epsfbox{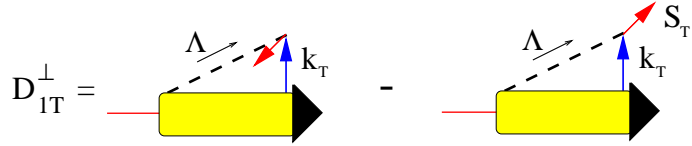}}
\caption{\label{D1Tperp}The chiral-even, T-odd 
function $D_{1T}^\perp$ signals different probabilities for $q \to
\Lambda(\bkt, \pm \bSt) + X$.}
\end{figure}

\begin{figure}[htb]
\epsfxsize=9cm \centerline{\epsfbox{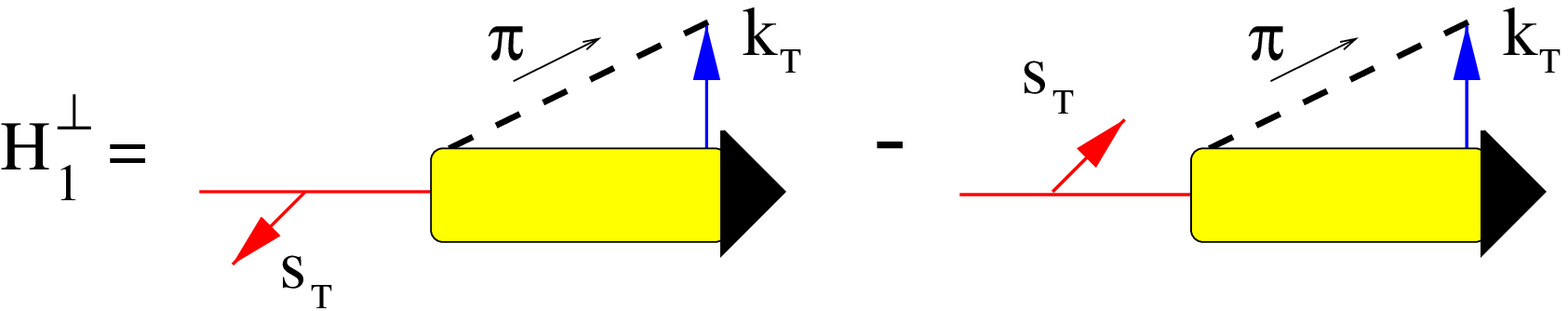}}
\caption{\label{H1perp}The Collins effect function
$H_1^\perp$ signals different probabilities for $q(\pm \bSt) \to
\pi(\bkt) + X$.}
\end{figure}

Therefore, in order to produce a large single transverse spin asymmetry in the
DY process (no experimental data exists however), one
needs some conceptually nontrivial mechanism, like soft gluon poles,
since regular perturbative and
higher twist contributions appear to be either small or absent. 
For the case of pion production in $p \, p^{\uparrow}$ scattering there exist
a more conventional explanation of the large observed single spin
asymmetries. It involves a T-odd {\em fragmentation\/} function. Such
functions are expected
to be present even if {time reversal invariance} is
applied, because of the final state interactions between the outgoing hadron 
and the other fragments.
For the description of a quark fragmenting into a pion plus anything, one
needs the fragmentation correlation function \cite{Coll-S-82} $\Delta(z)$,
which one parameterizes in accordance with 
the required symmetries (including time reversal)
\[
\Delta(z)=\text{T-even part} + \frac{M}{2} \left[ {D_T(z)} 
\epsilon_T^{\mu\nu} \gamma_{T\mu} S_{T\nu} - {E_L(z)} 
\lambda i\gamma_5 + {H(z)} \frac{i}{2} 
\left[\slsh{n_-},\slsh{n_+}\right] \right].
\]
The twist-3 fragmentation functions {$D_T,E_L,H$} are T-odd. 
The T-odd fragmentation function {$H$} can be responsible for the
single spin asymmetries, e.g.\ via a diagram as depicted in 
Fig.\ \ref{pppiX}.
This yields power suppressed asymmetries, which can be investigated 
experimentally 
by changing the energy scale over a wide range (again RHIC can provide this
information). The question we like to address here is: how to obtain a single 
spin asymmetry that is not suppressed by powers of the hard scale? 
A possible solution is to include intrinsic transverse momentum
\cite{Ralst-S-79}, i.e. replace  
$\Phi(x) \rightarrow \Phi(x,\bpt)$ and 
$\Delta(z) \rightarrow \Delta(z,\bkt)$. 
For the T-even distribution functions in DY this replacement 
leads to {\em double\/} 
spin asymmetries \cite{Ralst-S-79,Tangerman-Mulders-95a}.
But T-odd functions with transverse
momentum dependence can lead to {\em single\/} spin asymmetries {\em at leading
order\/}. 
The transverse momentum dependent distribution functions are defined as
\beq
\Phi(x,\bm{p}_T) 
=
\frac{1}{2}\,\biggl[
f_1\, \slsh{\! P}
- g_{1s}\, \slsh{\! P} \gamma_5
-h_{1T}\,i\sigma_{\mu\nu}\gamma_5 S_{T}^\mu P^\nu
- h_{1s}^\perp\,\frac{i\sigma_{\mu\nu}\gamma_5 p_T^\mu
P^\nu}{M} \biggl],
\label{paramPhixkt}
\eeq
where $f_1=f_1(x,\bpt)$, etc.\ and we 
use the shorthand notation
\beq
(..)_{1s}^{}(x, \bpt) \equiv
\lambda\,(..)_{1L}^{}(x ,\bm p_T^2)
+ \frac{(\bpt\cdot\bm{S}_{T})}{M}\,(..)_{1T}^{}(x ,\bm p_T^2).
\eeq
The correlation function $\Delta(z,\bkt)$ is given by
\beq
\Delta(z,\bkt)=
\text{T-even part} + \frac{1}{2} \biggl[
D_{1T}^\perp\, \frac{\epsilon_{\mu \nu \rho \sigma}
\gamma^\mu P^\nu k_T^\rho S_{T}^\sigma}{M}
+ H_{1}^\perp\,\frac{\sigma_{\mu \nu} k_T^\mu P^\nu}{M}
\biggr].
\label{Deltaexp}
\eeq
The fragmentation functions $D_{1T}^\perp$ and $H_1^\perp$ are T-odd
functions and as can be seen in Figs.\ \ref{D1Tperp} and \ref{H1perp} such 
T-odd effects link transverse momentum and 
transverse spin (orthogonal to the transverse momentum) 
with a specific orientation (handedness). 
The chiral-even function\cite{Mulders-Tangerman-96} $D_{1T}^\perp$ is
expected to be 
relevant for {transversely polarized $\Lambda$ production} \cite{ABM}, 
for instance in $p \, p \to \Lambda^\uparrow X$ (also measurable at RHIC). 
The chiral-odd function\cite{Collins-93b} $H_1^\perp$, also called 
the ``Collins effect'' function, might be relevant for 
$p \, p^\uparrow \to \pi \, X$ asymmetry via\cite{Anselmino}:
$A_T \sim h_1 (x_1)  \otimes f_1(x_2)  \otimes {H_1^\perp(z,\bkt)}$. 
Like the transverse
momentum of the lepton pair in the DY process, the transverse
momentum of the pion now originates from the intrinsic transverse momentum of
the initial partons in addition to transverse momentum generated 
perturbatively by radiating off some additional parton(s) in the final state 
(hence the transverse momentum of the pion need not be small). 

There is an experimental indication \cite{EST-98} 
from analyzing a particular angular dependence (a $\cos 2\phi$
dependence\cite{Boer2}) in the unpolarized process 
$e^+ \, e^- \rightarrow Z^0
\rightarrow \pi \, \pi \, X$, where the pions belong to opposite jets,
that the Collins effect is in fact a few percent of the magnitude of 
the ordinary unpolarized fragmentation function. Comparison to the
magnitude at lower energies of course requires evolution. 

The two T-odd fragmentation functions satisfy the following sum
rules\cite{Schafer-T-99}\footnote{Unlike Ref.\ \cite{Schafer-T-99} we use for 
the functions the notation {\em and\/} definitions of 
Ref.\ \cite{Mulders-Tangerman-96}}:
\beq
\sum_h \int dz \; z \; {H_1^{\perp
(1)}(z)}=  0,\qquad
\sum_h \int dz \; z \; {D_{1T}^{\perp (1) }(z)}= 0,
\eeq
where ${F^{(1)}(z)} = z^2 \int 
d^2 \bkt \, \bm{k}_T^2/(2M^2)\, F(z,z^2\, \bm{k}_T^2)$.

The function $D_{1T}^\perp$ can also be probed
in {charged current exchange processes}, since it is chiral-even, as opposed 
to the chiral-odd functions like $h_1, H_1^\perp$. 

So far we have not commented on the fact that the T-odd fragmentation fuctions
appearing in the parameterization of $\Delta(z)$ will lead to power suppressed
contributions, whereas the two T-odd functions of $\Delta(z,\bkt)$ can
contribute at leading order in the expansion in inverse powers of the hard
scale. This is the subject of the next section. 

\section{Transverse momentum at leading ``twist''}

A process like deep inelastic scattering has only two scales, the hadronic
scale (say the hadron mass $M$) and the large scale of the virtual photon 
($Q$). The
explicit mass term in front of the function $g_T$ in $\Phi(x)$ 
Eq.\ (\ref{paramPhiM})
can therefore only lead to a term $M/Q$ in the cross section. 

In the case of a less inclusive experiment, for instance $e \, p \to e' \, \pi
\, X$,
one can also observe the transverse momentum $\bm P_{\pi \perp}$ of the pion.
This semi-inclusive cross section
will depend on three dimensionful quantities: $M, |\bm P_{\pi \perp}|$ and 
$Q$. 
The function $g_T$ will again lead to a contribution $\sim M/Q$, since it is 
not sensitive to the transverse momentum of the pion; it will not average to 
zero if one averages over $\bm P_{\pi \perp}$. On the other hand, 
one might be sensitive to functions that will disappear upon averaging. For
these functions the appropriate scales are $M$ and $\bm P_{\pi \perp}$. 
Terms proportional to $M/|\bm P_{\pi \perp}|$ will appear 
{\em without\/} expanding in $M/|\bm P_{\pi \perp}|$. 
We will see an explicit example below. 

In order to arrive at a single transverse spin asymmetry that is not 
suppressed by inverse powers of the hard scale, one can consider cross
sections differential in the transverse momentum of the pion. In
that case one is sensitive to the transverse momentum of the quarks directly
and in case this concerns intrinsic transverse momentum of the quarks inside a
hadron, the effects need not be suppressed by $1/Q$. The point is that if the
transverse momentum of the pion is (solely) produced by perturbative QCD
corrections, each factor of transverse momentum has to be accompanied by the
inverse of the scale in the elementary hard scattering subprocess: $1/Q$. 
But in case of {\em intrinsic\/} transverse momentum of the quarks 
(or gluons) the relevant 
scale is not $Q$, but the hadronic scale $M$. 
In other words, one is not
allowed to make an expansion of $\Phi(x,\bpt)$ in terms of $\bpt/Q$, since
there could be terms of order $\bpt/M$, which are not small in
general. 

In processes with two
(or more) soft parts, like semi-inclusive leptoproduction, 
the intrinsic transverse
momentum of one soft part is linked to that of the other soft part 
resulting
in effects, e.g.\ azimuthal asymmetries, not suppressed by $1/Q$. These
effects will show up at relatively low (including nonperturbative) 
values of $|\bm P_{\pi \perp}|$.
By radiating off hard partons the fragmenting quark and hence the pion 
might achieve a higher transverse momentum. 
Of course, if this produces a very high 
transverse momentum ($\sim Q$), then this {\em will\/} lead to suppression. 
But the point is:
in case one observes the transverse momentum of the pion, one can probe
$\Phi(x,\bpt)$ and $\Delta(z,\bkt)$, without suppression by $1/Q$. 

Let us investigate the example\cite{Boer3} 
of $\vec e\, \vec p\rightarrow e \, \pi \, X$ 
and consider cross sections integrated, 
but weighted with a function of the transverse momentum of the pion:  
\beq
\langle W\rangle_{P_e P_p}
= \int d\phi^e\,d^2\bm P_{\pi\perp}^{}
\ W\,\frac{d\sigma_{P_e P_p}^{[\vec e \, \vec p\rightarrow e \, \pi \, X]}}
{dx\,dy\,dz\,d\phi^e\,d^2\bm P_{\pi\perp}^{}},
\label{asym}
\eeq
where $W$ = $W(|\bm P_{\pi\perp}^{}|,\phi_\pi^e,\phi_{S}^e)$.
$P_e$ and $P_p$ are the polarizations 
of the electron and proton with respect to the virtual photon.
We use $O$ for unpolarized, 
$L$ for longitudinally polarized ($\lambda \ne 0$) 
and $T$ for transversely polarized ($\vert \bm S_\st\vert \ne 0$) particles.

For pion production only expressions with the fragmentation functions
$D_1,H_1^\perp, E$ and $H$ contribute to the order we consider. 
The asymmetries involving a 
fragmentation function $D_1$ can also be obtained by looking at the
asymmetry in jet production. The jet is just a means of observing a 
transverse momentum. For simplicity we will now restrict to 
jet production (i.e., keeping only $D_1(z) = \delta(1-z)$) 
and consider pion production in the next section. 

For the case of $P_e P_p = L T$ 
we find the following power suppressed azimuthal spin asymmetry 
\beq
\frac{\langle 1
\rangle_{LT}}{\left[4\pi\,\alpha^2\,s/Q^4\right]} = -
\cos\phi_S^e \;
\lambda_e\,\vert\bm S_\st\vert
\,y\,\sqrt{1-y} \ \frac{M}{Q} \, \sum_{a,\bar a} e_a^2\, 
x^2\,g_T^a(x).
\eeq
On the other hand, if one weights with powers of the observed 
transverse momentum (hence $\langle |\bm P_{\text{jet}\, \perp}| \rangle$ is a
scale in the problem) one obtains for instance the following leading order
expression 
\cite{Boer3}
\beq
\frac{\big\langle 
(|\bm P_{\text{jet} \,\perp}| /M)\,\cos \phi_{\text{jet}}^e 
\big\rangle_{LT}}{\left[4\pi\,\alpha^2\,s/Q^4\right]} = \cos \phi_S^e \; 
\lambda_e\,\vert \bm S_\st\vert\,y\,(1-\frac{1}{2}\,y) \sum_{a,\bar a} e_a^2
\,x\,g_{1T}^{(1)a}(x).
\label{WLT}
\eeq
The function $g_{1T}^{(1)a}(x) = \int 
d^2 \bpt \, \bm{p}_T^2/(2M^2) \, g_{1T}^{a}(x,\bm{p}_T^2)$ 
appears in the quantity 
\beq
\Phi_\partial^\alpha(x) \equiv \int d^2 \bm{p}_T \, p_T^\alpha \,  
\Phi (x,\bpt)  = 
\frac{M}{2} 
\bigg[ g_{1T}^{(1)} (x)\,
S_{T}^\alpha \gamma_{5} \mbox{$\not\! P\,$} - 
h_{1L}^{\perp (1)}(x) \lambda\gamma_{5} \gamma_T^\alpha \mbox{$\not\! 
P\,$}
\bigg].
\label{Phipartial}
\eeq
Clearly the factor $M$ has to be compensated in the cross section and as is
seen from Eq.\ (\ref{WLT}) the relevant scale is $1/\langle 
|\bm P_{\text{jet} \,\perp}| \rangle$. 
Nevertheless, from the QCD e.o.m.\ it is clear that 
$\Phi_\partial^\alpha(x)$ is related to a quark-gluon-quark matrix element
that will always show up $M/Q$ suppressed {\em in DIS}. And in fact, 
the function $g_{1T}^{(1)}$ is a well-known quantity in the Wandzura-Wilczek
approximation: 
it equals (upon neglecting quark
masses) $x \, g_T^{WW}(x)$, where $g_T^{WW}= g_1 + g_2^{WW} $. 
The existing data on $g_T$
are (still) consistent with 
$g_T= g_T^{WW}$. 
This shows that part of the ``twist-3''
information (the dominant contribution in fact) can be obtained without
suppression factors of $1/Q$, by doing a less inclusive experiment, namely by
observing a transverse momentum. 
Of course the average transverse momentum
is also a function of $Q$ and experimentally the observation of transverse
momenta is more difficult, but the point remains that the
same information (e.g.\ $g_T^{WW}$) can enter with
different scales in different quantities. 

\section{Single spin asymmetry in \protect{$e \, \vec{p} \to e' \, \pi^+ \,
X$}}

\begin{figure}[htb]
\epsfxsize=9cm \centerline{\epsfbox{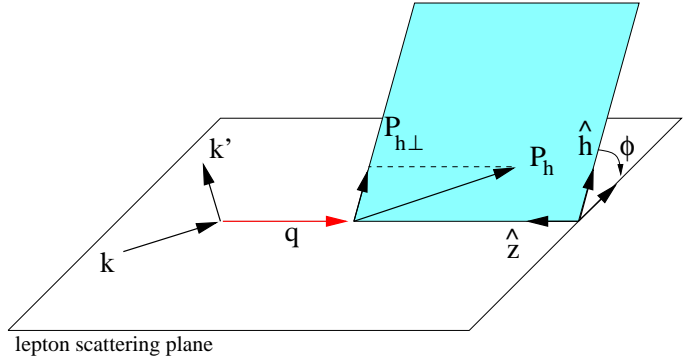}}
\caption{\label{kindis}Kinematics of the $\sin \phi$ asymmetry in 
semi-inclusive deep inelastic scattering.}
\end{figure}

Recently, the HERMES Collaboration \cite{HERMES} reported a $\sin \phi$ 
asymmetry in the process $e \, \vec{p} \to e' \, \pi^+ \, X$,
where the target has a polarization along
the electron beam direction and $\phi \, (= \phi^e_\pi)$ 
is the angle of the transverse
momentum of the pion with respect to the lepton scattering plane, cf.\ Fig.\ 
\ref{kindis}.
The measured asymmetry has an analyzing power of 
$0.022 \pm 0.005 \pm 0.003$. The fact that the pion prefers to be out-of-plane
in the asymmetry (a $\sin \phi$ distribution) is indicative of a T-odd
effect. The asymmetry can indeed be expressed in terms of a 
chiral-odd, T-odd fragmentation function with transverse momentum dependence, 
the Collins effect function $H_1^\perp$, if a factorized 
picture like in Ref.\ \cite{Mulders-Tangerman-96} is assumed:
\beq
{\cal W}^{\mu\nu}= \int d^2\bm{p}_T^{} d^2 
\bm{k}_T^{}\, \delta^2(\bm{p}_T^{}+
\bm{q}_T^{}-\bm{k}_T^{})\, 
\text{Tr}\bigl[ {\Phi (x,\bpt)} 
\gamma^\mu {\Delta 
(z,\bkt)} \gamma^\nu 
\bigr].
\eeq
One can easily project out the $\sin \phi$ dependence from the 
cross section by weighted integration. We define $A_{P_e P_p} 
= \langle W\rangle_{P_e P_p}/ \langle 1\rangle_{OO}$ with 
$W=(|\bm{P}_{\pi\perp}|/zM_\pi)\;\sin \phi^e_\pi$.

We will assume the asymmetry arises mainly from the dominant flavor, i.e.\ we
take into account only the contribution from $u \to \pi^+$. Furthermore, we
will neglect transverse momentum effects inside the proton.
We find the following expressions for the relevant asymmetries:
\ba
A_{{OL}} & \propto & \lambda \; 2 (2-y)\sqrt{1-y}\;
{\frac{M}{Q}}
\; {x \; h_L(x)} \; {H_1^{\perp (1)}(z)},\\
A_{{OT}}& \propto & |\bm S_{\st}^{}|\; (1-y)\;{h_1(x)} 
\; {H_1^{\perp (1)}(z)},
\ea 
where ${H_1^{\perp (1)}(z)} = z^2 \int 
d^2 \bkt \, \bm{k}_T^2/(2M_\pi^2) \, H_1^{\perp}(z,z^2\, \bm{k}_T^2)$. 
 
\begin{figure}[htb]
    \parbox{\halftext}{
                \epsfxsize=6.6cm \centerline{\epsfbox{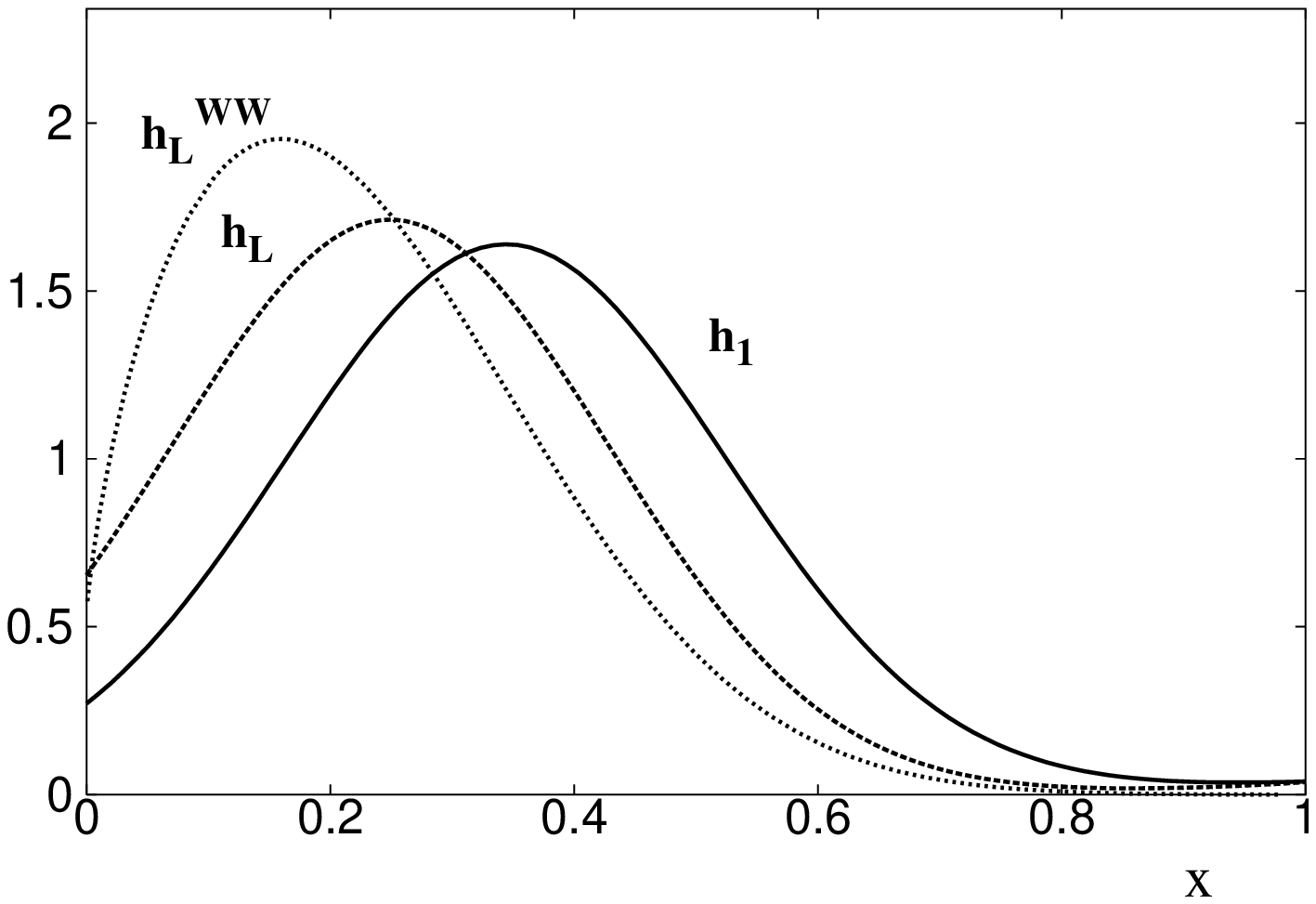}}
                \caption{\label{Bag1}Bag model functions $h_1$, $h_L$ and
                $h_L^{WW}$.}}
            \hspace{6mm}
            \parbox{\halftext}{
                \epsfxsize=6.6cm \centerline{\epsfbox{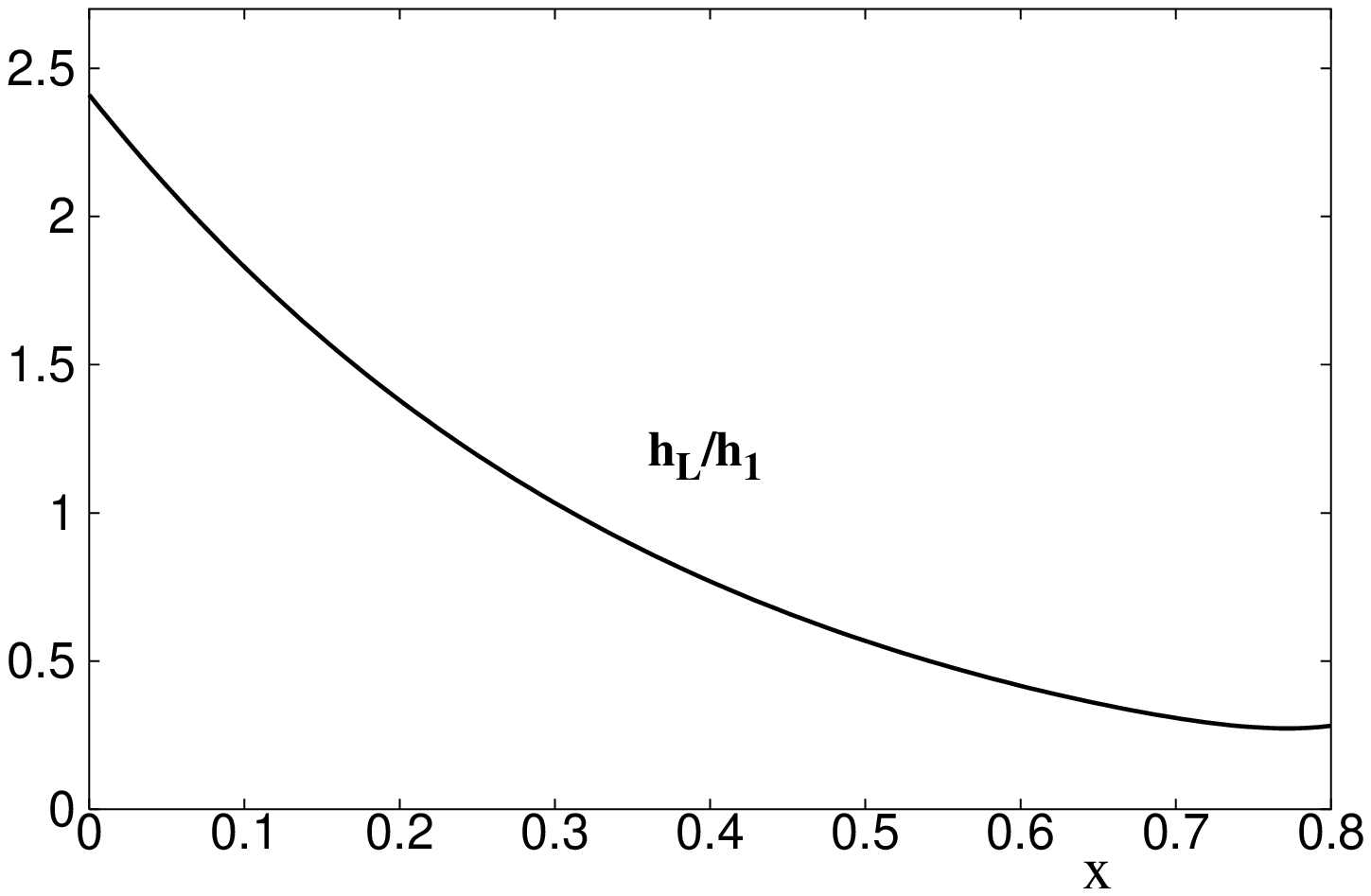}}
		\caption{\label{Bag2}Bag model ratio $h_L/h_1$.}}
\end{figure}

The polarization of the target $P_p$ 
is in the lepton scattering plane and in fact 
along the electron beam direction, hence, it is 
a combination of $L$ and $T$, depending on $y= (P\cdot q) / (P\cdot
l)$. We find
\beq
\frac{|\bm S_{\st}^{}|}{\lambda}= \sqrt{1-y}\;  
\frac{2{M}x}{{Q}}\; \quad \Longrightarrow \quad  
\frac{A_{O{L}}}{A_{O{T}}}= \frac{2-y}{1-y} \; 
{\frac {h_L(x)}{h_1(x)}}.  
\label{Splam}
\eeq
A striking feature is that the 
contribution from the target spin {transverse} to the virtual photon 
momentum also enters at subleading order in {$M/Q$}, even though 
the relevant functions ($h_1$ and $H_1^\perp$) are leading twist
functions. 
For $0.2 < y < 0.8$ one then finds
\beq
2\; \frac{h_L}{h_1} \, \simordertwo \, \frac{A_{OL}}{A_{OT}} \, \simordertwo \,
6 \; \frac{h_L}{h_1}.
\eeq
Using a bag model calculation \cite{Jaffe-Ji-91}
we confirm the
observation of Kotzinian {\em et al.} \cite{Kotzinian}, that $A_{O{L}}$ 
is the dominant contribution to the asymmetry (for $0 < x < 0.3$), see Figs.\
\ref{Bag1} and \ref{Bag2}. 
In Ref.\ \cite{Kotzinian} they consider the Wandzura-Wilczek approximation; 
the bag 
model calculation of $h_L^{WW} = - 2 \, h_{1L}^{(1)\, WW}/x$ deviates 
considerably from $h_L$ in the small $x$ region though. 
We also observe that for larger $x$ (above $0.5$), but 
smaller values of $y$, 
$A_{OT} \simorder A_{OL}$ (we restrict to  
$x < 0.8$ due to model artifacts).  

Measuring the same asymmetry for a target
polarized {\em transversely\/} 
to the electron beam, really probes the unsuppressed
asymmetry $A_{OT}$, since in this case $A_{OL}$ will contribute $M^2/Q^2$ 
suppressed as one can see
by interchanging the role of $|\bm S_{\st}^{}|$ and $\lambda$ in 
Eq.\ (\ref{Splam}).  

The asymmetry $A_{{LO}}$ has also been
measured and was found to be consistent with zero, coinciding with the
Wandzura-Wilczek expectation:
\ba
A_{{LO}} & \propto &\lambda_e \; 2y \sqrt{1-y} \; {\frac{M}{Q}}\;
{x\; e(x)} 
\;{H_1^{\perp (1)}(z)} \stackrel{WW}{\propto} \frac{m_u}{Q} \approx 0.
\ea

We conclude that the HERMES Collaboration might have obtained the first 
experimental information on the functions $h_1$ and $h_L$. 

\vspace{-1 mm}
\section*{Acknowledgements}
We thank M.~Anselmino, A.V.~Belitsky, A.V.~Efremov, R.L.~Jaffe, 
R.~Jakob, P.J.~Mulders, K.A.~Oganessyan, N.~Saito for valuable 
discussions. Furthermore, we thank RIKEN, 
Brookhaven National Laboratory and the U.S.\ 
Department of Energy (contract number DE-AC02-98CH10886) for
providing the facilities essential for the completion of this work.

\end{document}